# pLMFPPred: a novel approach for accurate prediction of functional peptides integrating embedding from pre-trained protein language model and imbalanced learning

Zebin Ma[a], Yonglin Zou[a], Xiaobin Huang[a], Wenjin Yan[b], Hao Xu[a], Jiexin Yang[a], Ying Zhang[a, *], Jinqi Huang[c, *]

a. School of Mathematics and Computer Science, Guangdong Ocean University, Guangdong, 524088, China.

b. The Institute of Pharmacology, Key Laboratory of Preclinical Study for New Drugs of Gansu Province, School of Basic Medical Sciences, Lanzhou University, Gansu, 730000, China.

c. Department of Hematology, Affiliated Hospital of Guangdong Medical University, Guangdong, 524000, China.

# Authors

Zebin Ma; Email: mazebin@stu.gdou.edu.cn; School of Mathematics and Computer Science, Guangdong Ocean University, Guangdong, 524088, China.

Yonglin Zou; Email: zouyonglin@stu.gdou.edu.cn; School of Mathematics and Computer Science, Guangdong Ocean University, Guangdong, 524088, China.

Xiaobin Huang; Email: huangxiaobin@stu.gdou.edu.cn; School of Mathematics and Computer Science, Guangdong Ocean University, Guangdong, 524088, China.

Wenjin Yan; Email: yanwj@lzu.edu.cn; The Institute of Pharmacology, Key

Laboratory of Preclinical Study for New Drugs of Gansu Province, School of Basic Medical Sciences, Lanzhou University, Gansu, 730000, China.

Hao Xu; Email: 202113211232@stu.gdou.edu.cn; School of Mathematics and Computer Science, Guangdong Ocean University, Guangdong, 524088, China.

Jiexin Yang; Email: 202113211432@stu.gdou.edu.cn; School of Mathematics and Computer Science, Guangdong Ocean University, Guangdong, 524088, China.

# Corresponding Authors

* Ying Zhang; Email: zhangying@gdou.edu.cn; Tel: +86 13828205545; School of Mathematics and Computer Science, Guangdong Ocean University, Guangdong, 524088, China. Handling correspondence at all stages of refereeing and publication, also post-publication.

* Jinqi Huang; E-mail: Jinqi@gdmu.edu.cn; Tel: +86 0759-2387411; Department of Hematology, Affiliated Hospital of Guangdong Medical University, Guangdong, 524000, China.

# Abstract

## Background

Functional peptides have the potential to treat a variety of diseases. Their good therapeutic efficacy and low toxicity make them ideal therapeutic agents. Artificial intelligence-based computational strategies can help quickly identify new functional peptides from collections of protein sequences and discover their different functions.

## Results

Using protein language model-based embeddings (ESM-2), we developed a tool called pLMFPPred (Protein Language Model-based Functional Peptide Predictor) for predicting functional peptides and identifying toxic peptides. We also introduced SMOTE-TOMEK data synthesis sampling and Shapley value-based feature selection techniques to relieve data imbalance issues and reduce computational costs. On a validated independent test set, pLMFPPred achieved accuracy, Area under the curve - Receiver Operating Characteristics, and F1-Score values of 0.974, 0.99, and 0.974, respectively. Comparative experiments show that pLMFPPred outperforms current methods for predicting functional peptides.

## Conclusions

The experimental results suggest that the proposed method (pLMFPPred) can provide better performance in terms of Accuracy, Area under the curve - Receiver Operating Characteristics, and F1-Score than existing methods. pLMFPPred has achieved good performance in predicting functional peptides and represents a new computational method for predicting functional peptides. The source code and dataset can be obtained at https://github.com/Mnb66/pLMFPPred.

# Keywords

# Background

Recently, peptides with various functional properties have been discovered, and these properties offer novel ideas for treating human diseases. Compared to their small-molecule counterparts, peptides have more H-bonded donors and receptors, suggesting they could bind to their targets with exquisite specificity and relatively few off-target side effects [1, 2]. At the same time, peptides are readily degradable by enzymes and have low toxicity associated with peptide metabolism. With the possibility of both probing and modulating protein−protein interactions, peptides are considered ideal therapeutics [3].

Computer-aided methods based on machine learning strategies provide fast and effective proteomic and synthetic sequence screening alternatives to accelerate the discovery of novel functional peptides. For example, antifp [4] combines peptide composition features and support vector machines to predict antifungal peptides; MAHTPred [5] predicts anti-hypertensive peptides based on feature descriptors and integrated learning, effectively improving balanced prediction performance and model robustness on independent datasets; and AntiCP2.0 [6] uses an extreme random tree classifier model for anti-cancer peptides. In addition to machine learning methods, deep learning methods have also been used to identify peptide properties. The formation of natural proteins or peptides can be analogous to natural language, which allows deep learning to decipher the information in peptide sequences directly and accurately [7]. For example, ATSE [8] predicts peptide toxicity based on graph neural networks and attention mechanisms; ToxIBTL [9] uses migration learning to build a deep learning framework for predicting peptide toxicity; AFPDeep [10] constructs a hybrid neural network model with convolutional neural network (CNN) layers and long short-term memory (LSTM) layers intertwined to predict antifungal peptides; Deep- AntiFP [11] used a deep neural network tuned with hyperparameters for antifungal peptide prediction; Deep-AFPpred [12] used migration learning and a 1DCNN-BiLSTM neural network for antifungal peptide classification to improve the classification accuracy of antifungal peptides; AniAMPpred [13] combined deep learning-based features with support vector machine for predicting potential antimicrobial proteins; Deep-ABPpred [14] constructed a bidirectional long short-term memory network based on word2vec amino acid features for identifying antimicrobial peptides; and a transformer-based approach [15] was proposed to extract peptide sequence information using the transformer architecture and natural language processing knowledge to identify antimicrobial peptides and their functional activities;

Language models (LMs) have recently emerged as a powerful paradigm for learning embeddings directly from large, unlabelled natural language datasets. These advances are now being explored in protein research through protein language models (pLMs) [16]. pLMs are trained by masking some part of the input protein sequence (usually a single amino acid) and reconstructing it from the uncorrupted sequence context. During inference, pLMs use the output of the last hidden layer of the network to represent a numerical vector of protein sequences, called an embedding. This allows information gathered from large but untagged protein sequence databases to be transferred to much smaller but tagged sequence datasets. To make the most of the vast amount of potential information in large protein sequence databases, various pLMs [17, 18, 19, 20, 21] have been developed to extract information from them [22] and transfer the information to other tasks.

Class imbalance, depicted as an unequal distribution of known classes in a dataset under classification problems, with one or more classes having significantly fewer samples than others. This can lead to a model's predictions being biased towards the majority class and ignoring the minority class, resulting in poor performance [23]. Various algorithms have been proposed to relieve the class imbalance problem [24, 25]. The synthetic minority oversampling technique (SMOTE) [26] is one of the most classical and effective methods that have been used in bioinformatics research [27].

In this paper, a new deep learning-based method called pLMFPPred is proposed. This method utilizes knowledge transferred from a large protein sequence database using the pLM ESM-2 model to extract embeddings to improve the recognition and prediction of three types of functional peptides and toxic peptides. The method also employs an imbalanced learning strategy to relieve class imbalance. pLMFPPred achieves performance that meets or exceeds that of existing methods on benchmark datasets. The performance of different embeddings and feature encodings was also compared. Experimental results show that transfer learning and imbalanced learning can be integrated to improve biological sequence prediction effectively.

# Material and Methods

## Data preprocessing

It is widely known that the quality and quantity of datasets play a crucial role in the performance of a model. In this paper, we will focus on three different types of functional peptides (antibacterial, antifungal, and antihypertensive) and peptides with toxicity, using datasets validated by previous related work. Specifically, the antifungal dataset was collected from antifp [4], the antihypertensive dataset from MAHTPred [5], the antibacterial dataset from Deep-ABPpred [14], and the toxic peptide dataset from ATSE [8]. The quantity of each data set can be found in Table 1.

We performed the preparation of the dataset in two steps. First, we remove duplicate entries using CD-HIT [28] on sequences under each target domain independently with a threshold of 40%. Then, if any duplicate sequences existed between the functional peptide data and the toxic peptide data, the duplicates were retained in the toxicity dataset. We ended up with 10019 sequences containing 1921 toxic peptides, 5619 antimicrobial peptides, 1052 antihypertensive peptides, and 1427 antifungal peptides. We then used a fixed random seed to split the data into an 80% training set (8015) and a 20% test set (2004). Figure 1 shows the flowchart of data preprocessing.

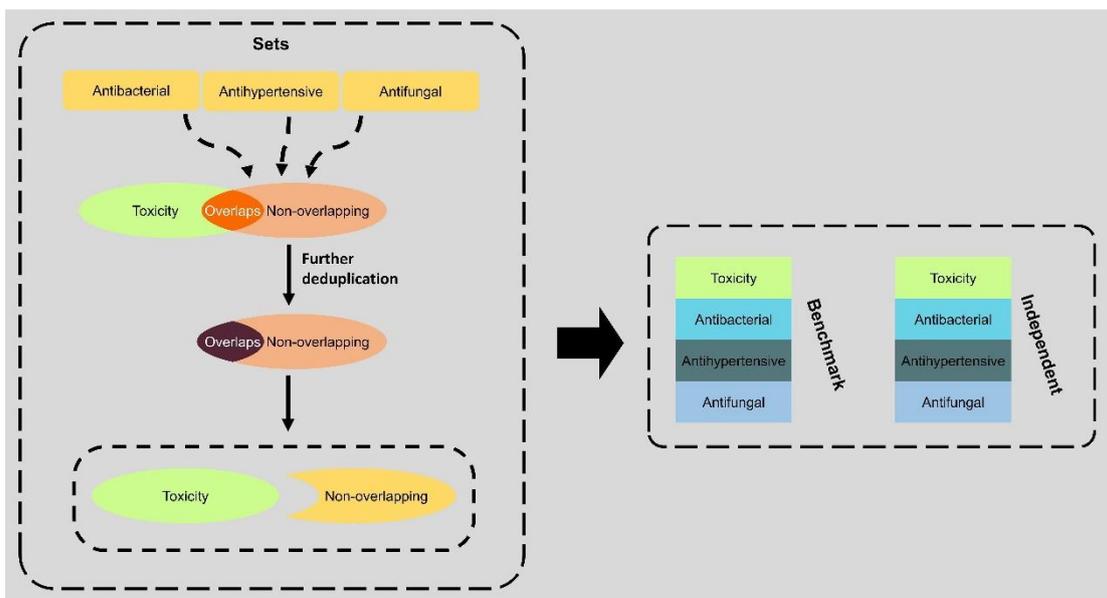

Fig 1. Flowchart of splitting the dataset.

Table 1. Summary of Data Sets.

| Category | Benchmark | Independent |
| --- | --- | --- |
| Toxicity | 1535 | 386 |
| Antifungal | 1131 | 296 |
| Antihypertensive | 856 | 196 |
| Antibacterial | 4493 | 1126 |
| Total | 8015 | 2004 |

# Embedding from pre-trained protein language model ESM-2

In this paper, the pLM from ESM-2 [29] was used in pLMFPPred. We used the embedding obtained from it to improve the prediction. The ESM-2 model uses a self-attention mechanism [30] to automatically capture the intrinsic relationships between all possible amino acid pairs within the input sequence to improve the representation.

ESM-2 is based on the UniRef50 dataset from The UniProt Reference Clusters [31] for self-supervised training. Unlike supervised word embedding, the input to the model is a full-length protein sequence, and the model predicts the identity of randomly selected amino acids in the protein sequence by looking at their context. This prompts the model to learn the association relationships between amino acids. During pre-training, a portion of the input markers are randomly masked. Accordingly, the model uses the untagged sequence data through a pre-training procedure to gain relatively complete knowledge of the amino acid sequences, which is transferred to downstream tasks to improve prediction performance. The architecture of the ESM-2 model consists of 48 hidden layers, each consisting of 40 self-attentive heads, with approximately 15 billion learnable parameters.

In pLMFPPred, The pre-trained ESM-2 model takes the entire protein sequence as input and returns a 5120-dimensional embedding vector for each amino acid. To complete the classification task, the embedding output is averaged and fed downstream to Light Gradient Boosting [32] for prediction. Figure 2 shows the flowchart of pLMFPPred.

## Class imbalance learning strategy based on SMOTE-TOMEK

Class imbalance is the most common problem in classification tasks. Figure 3 shows the proportion of labels for different kinds of functional peptides in the dataset, indicating the presence of imbalance in the data. To relieve the class imbalance problem in the dataset, we introduced the SMOTE-TOMEK data synthesis sampling method [33]. The method first uses the SMOTE method to generate synthetic samples and then uses the Tomek links algorithm to identify and remove noisy samples. It is important to note that data sampling is only performed on the training set to avoid introducing bias into the performance evaluation of the model.

## Feature selection

The embedding features extracted by ESM-2 have 5120 dimensions, and such high-dimensional data are prone to introduce interference noise and consume a large amount of resources during computation. To reduce interference noise between embeddings, improve training speed and reduce the computational cost, we introduced a feature selection method based on statistical hypothesis testing and efficacy calculation while combining Shapley values [34] to select the best feature subset and remove features that do not contribute or have a negative impact to the model.

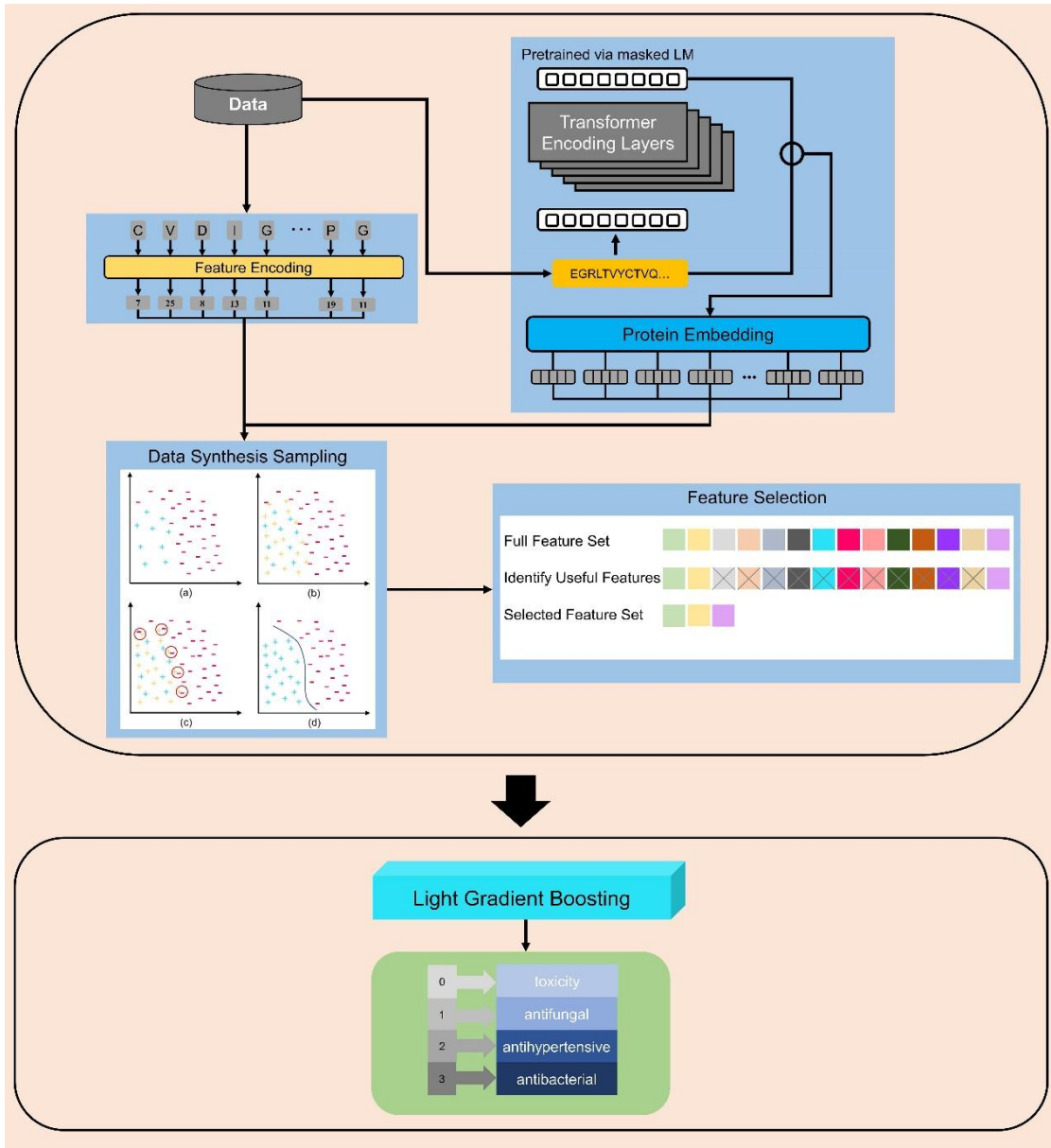

Fig 2. Flowchart of the pLMFPPred.

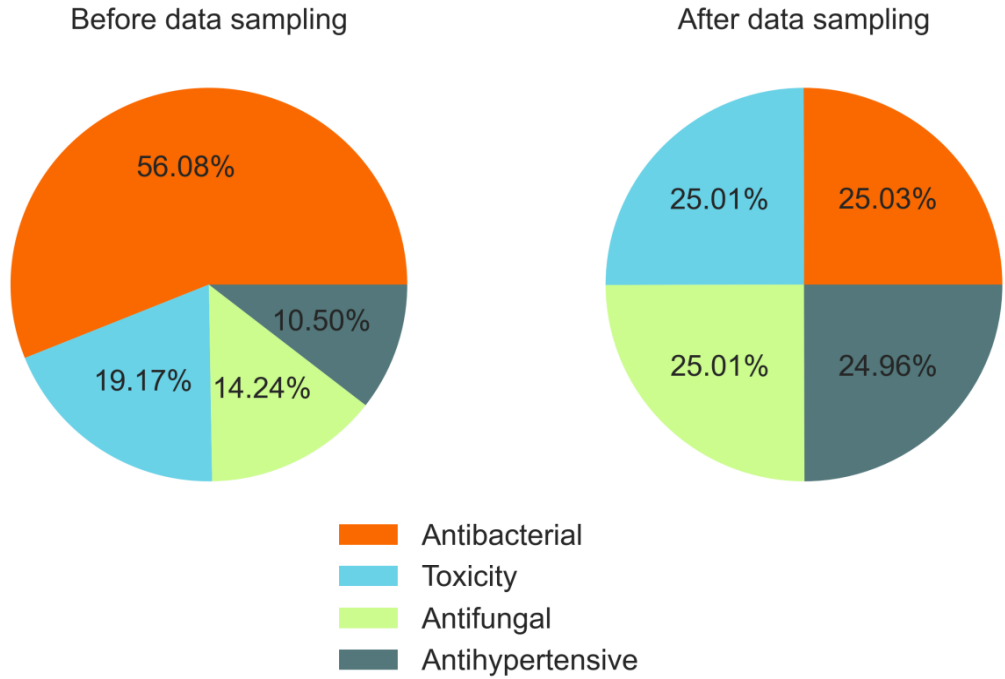

Fig 3. The distribution of the data set before and after data sampling, indicating the existence of imbalance in the original data, while the imbalance between data categories largely disappears after performing SMOTE-TOMEK data sampling.

## Performance metrics

Below, we define the performance metrics used for evaluating the models. We used Area under the curve - Receiver Operating Characteristics (AUC-ROC), accuracy, precision, recall, and F1-Score to evaluate the performance of the model.

$$False\ Positive\ Rate (FPR) = \frac{FP}{TN + FP}$$

$$True\ Positive\ Rate (TPR) = recall = \frac{TP}{TP + FN}$$

$$precision = \frac{TP}{TP + FP}$$

$$accuracy = \frac{TP + FN}{TP + FN + TN + FP}$$

$$F1\_score = \frac{2 \times precision \times recall}{precision + recall}$$

Where TP = True Positive, FP = False Positive. TN = True Negative, FN = False Negative.

## Comparisons of model performance

To compare with other methods, four sequence descriptors and embeddings extracted using two popular protein language models were used for comparison with the ESM-2 based pLMFPPred. The sequence descriptors included the composition-transition-distribution transition (CTDT), grouped amino acid composition (GAAC), pseudo amino acid composition (PAAC), and amino acid composition (AAC); four sequence descriptors were extracted using iFeature [35]. The protein language models included a pre-trained BERT model called TAPE [18] and ProtT5 [36]. We trained three ML-based models to compare downstream predictors, including eXtreme Gradient Boosting, Random Forest, and Support Vector Machine (SVM). We also trained a Bi-directional Long Short-Term Memory (BiLSTM) network for comparison with popular deep learning-based neural network models. We also trained another pLMFPPred trained with the original training set without imbalance learning for comparison to assess the effectiveness of the imbalance learning strategy.

# Results

## Performance of base models and pLMFPPred on the independent test set

As shown in Table 2, pLMFPPred achieved the best accuracy (0.974), and the best F1-score (0.974) of all models, followed by XGboost (ACC=0.973, F1-Score=0.973) and the Bilstm model (ACC=0.966, F1-Score=0.966), with random Forest and SVM performed moderately. The information gain obtained from the computational embedding of the ESM-2 model allowed pLMFPPred to achieve the best performance of all models. The performance of pLMFPPred using imbalanced learning improved compared to pLMFPPred trained with class-imbalanced sequence data. This suggests that the imbalance learning strategy reduces the degree of imbalance in the different classes of data, allowing the model to more accurately classify the data by overcoming the bias that exists between the imbalanced classes, resulting in a more balanced and robust model performance. Figure 4 shows the AUC-ROC curves for these categories. Each category has an impressive AUC value of 0.99 or above. Table 3 shows the feature selection results, with the model feature dimension reduced from 5120 to 340, a 93.4% reduction in feature dimensionality, and a 92.6% reduction in training time from 74s to 5.5s. In contrast, pLMFPPred showed only a 0.2%-0.3% performance loss. This indicates that the feature selection effectively removes features that could lead to overfitting or noise, significantly reducing the dimensionality of the dataset. The reduction in model training time also implies a significant reduction in the consumption of computational resources. We believe that a performance loss of 0.2%-0.3% is acceptable in this case. It is worth noting that the neural network-based deep learning classifier did not outperform pLMFPPred with Light Gradient Boosting as the classifier. This suggests that the advantage of the deep learning-based neural network classifier may be insignificant in this type of task. The above results demonstrate the excellent and stable performance of pLMFPPred.

Table 2. pLMFPPred's performance compared to existing methods. The best results are shown in bold, and the second-best results are underlined.

|  | Precision | Recall | F1-Score | Accuracy |
|---|---|---|---|---|
| pLMFPPred | **0.974** | **0.974** | **0.974** | **0.974** |
| pLMFPPred w/o imbalanced learning | <u>0.973</u> | 0.954 | 0.963 | <u>0.973</u> |
| BiLSTM | 0.966 | <u>0.966</u> | <u>0.965</u> | 0.966 |
| XGBoost | 0.967 | 0.955 | 0.961 | <u>0.973</u> |
| Random Forest | 0.937 | 0.883 | 0.907 | 0.929 |
| SVM | 0.884 | 0.783 | 0.825 | 0.868 |

Table 3. pLMFPPred's performance before and after feature selection.

|  | Feature Dimension | Training Time | Precision | Recall | F1-Score | Accuracy |
|---|---|---|---|---|---|---|
| Before Feature | 5120 | 74s | 0.974 | 0.974 | 0.974 | 0.974 |

| Selection | | | | | | |
|---|---|---|---|---|---|---|
| After Feature Selection | 340 | 5.5s | 0.972 | 0.972 | 0.971 | 0.972 |

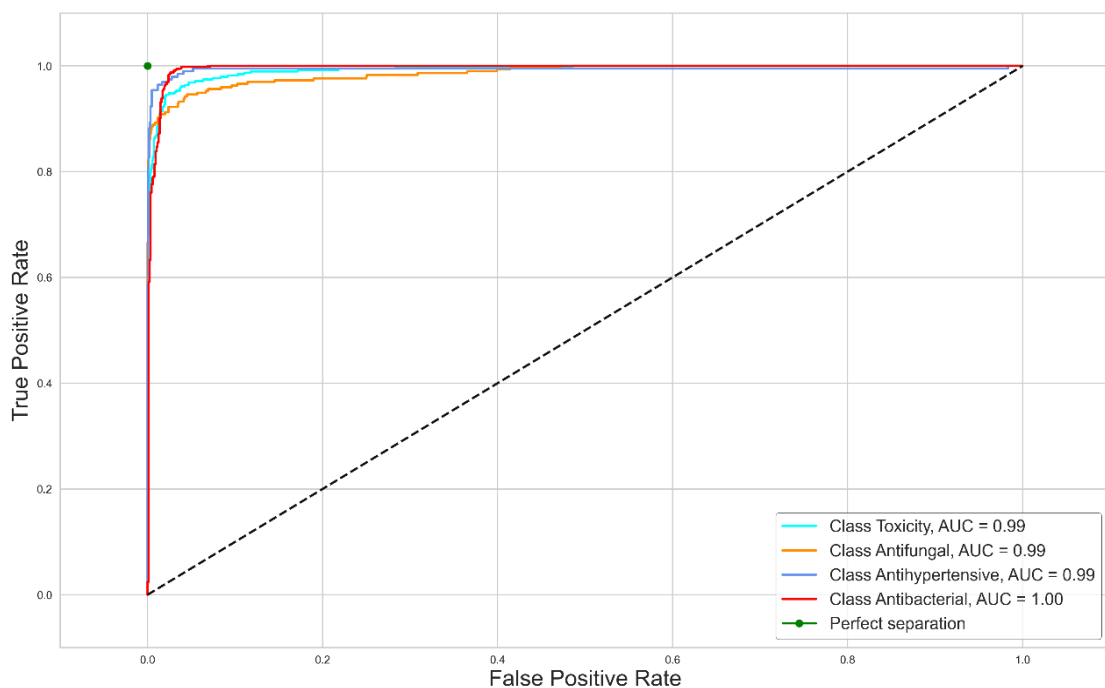

Fig 4. The AUC-ROC curve of pLMFPPred.

# Performance comparison of different embedding and feature coding

Table 4 summarises the performance of each model, with the ESM-2 model achieving the best accuracy of 96.4% and the best F1-Score of 96.3%, outperforming ProtT5 and TAPE, while the traditional feature descriptors performed moderately. The results show that the ESM-2 based pLMFPPred achieves the best performance. We also tested ESM-2 models with different parameter scales, and as shown in Table 5, the ESM-2 model with 15 billion parameters outperformed all other models. This suggests that as the parameter scale of the model rises, the protein language model's ability to understand the sequence improves, leading to better performance in downstream prediction tasks as well.

Table 4. Comparison of performance for different embeddings and descriptors. Best results are bold, and the second-best results are underlined.

|  | Precision | Recall | F1-Score | Accuracy |
|---|---|---|---|---|
| ESM-2 | **0.964** | **0.964** | **0.963** | **0.964** |
| ProtT5 | <u>0.963</u> | <u>0.963</u> | <u>0.963</u> | <u>0.963</u> |
| TAPE | 0.962 | 0.962 | 0.961 | 0.962 |
| CTDT | 0.916 | 0.917 | 0.915 | 0.917 |
| AAC | 0.950 | 0.949 | 0.949 | 0.949 |
| GAAC | 0.874 | 0.878 | 0.875 | 0.878 |
| PAAC | 0.953 | 0.953 | 0.953 | 0.953 |

Table 5. Comparison of performance for ESM-2 models with different parameter scales. Best results are bold, and the second-best results are underlined.

| ESM-2 Parameter scale | Precision | Recall | F1-Score | Accuracy |
|---|---|---|---|---|
| 35M | 0.951 | 0.951 | 0.950 | 0.951 |
| 150M | 0.958 | 0.958 | 0.958 | 0.958 |
| 650M | <u>0.960</u> | <u>0.960</u> | <u>0.959</u> | <u>0.960</u> |
| 3B | 0.957 | 0.957 | 0.957 | 0.957 |
| 15B | **0.964** | **0.964** | **0.963** | **0.964** |

## UMAP visualization of pLMFPPred

Compared to the baseline model, the ESM-2 based pLMFPPred exhibits considerable performance improvements. The main reason for these improvements can be attributed to the latent features learned and extracted by ESM-2 from the peptide sequences. To visually demonstrate the effectiveness of the embedding features extracted by ESM-2, we compared the ESM-2 embedding encoded features with the baseline dataset features by employing Uniform Manifold Approximation and Projection for Dimension Reduction (UMAP) [37] to present a clear representation of the features. Figure 5 shows the results of the visualization. Compared to the baseline dataset features, the ESM-2 embedding encoded features showed a clear impact on facilitating the intrinsic separation of various classes of functional peptides, with the features learned from the embedding shown to enable the formation of distinct clusters between peptides with different functional targets, with the boundaries of this separation being more apparent compared to the baseline feature encoding. Feature encoding from the baseline approach fails to achieve differentiation in the UMAP representation, which manifests as many peptides with different functions becoming entangled together. This analysis suggests that pLMFPPred based on the ESM-2 embedding encoding feature can produce a more accurate representation to help differentiate between peptides with different functions.

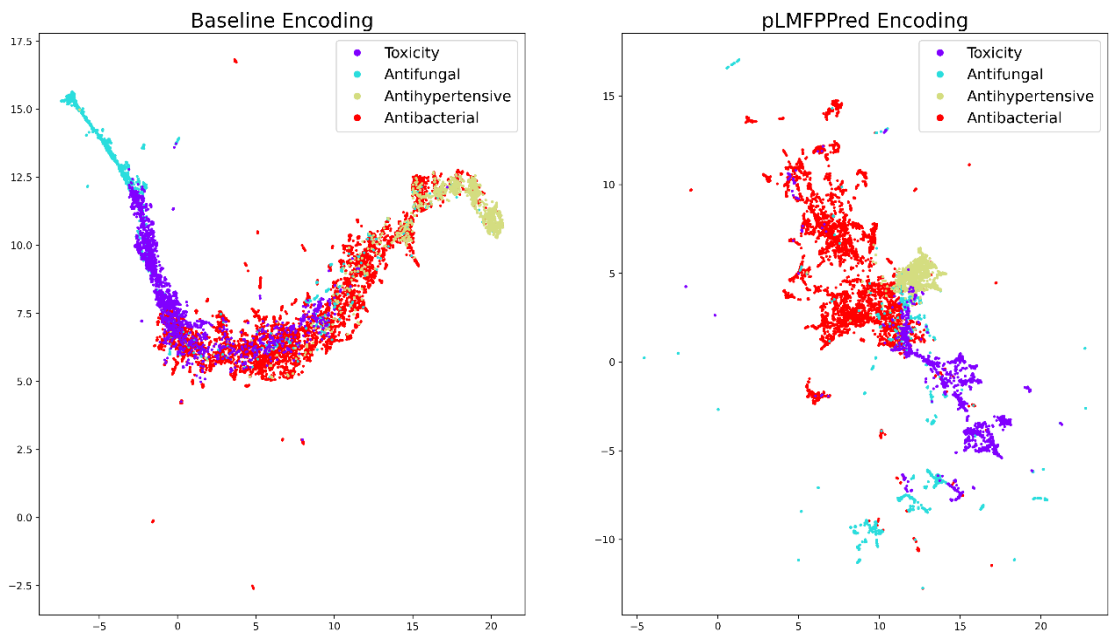

Fig 5. UMAP visualization of sequences for pLMFPPred and the baseline encoding.

# Discussion and conclusions

In this paper, we present a novel computational approach called pLMFPPred, a functional peptide prediction tool using pre-trained protein language model. The proposed method combines embedded features extracted by ESM-2 with SMOTE-TOMEK data synthesis sampling and feature selection to relieve data imbalance and reduce computational costs. We used the model to identify and predict three different Functional Peptides and peptides with toxicity. Experimental results show that pLMFPPred outperforms other state-of-the-art methods in various metrics such as AUC-ROC, Accuracy, and F1-Score, confirming the proposed model's ability to address data imbalance, reduce computational cost, and improve Functional Peptide prediction performance. We believe that the proposed scheme combining protein language model with feature selection and imbalanced learning techniques can be widely applied to other biological sequence analysis problems.

The limitation of this study is that the scale of the functional peptide dataset used still needs to be improved. In the future, we will fully utilize the rich sequence data from large protein sequence databases [38] to expand the functional peptide repertoire to effectively differentiate a wider range of functional peptides.

In addition, although the computational cost of downstream prediction tasks is effectively reduced by using additive models based on tree ensembles (rather than complex neural networks) and feature selection while ensuring good performance metrics, appropriate computational resources are still necessary due to the use of large protein language models involved. Therefore, the development of more lightweight pLMs is a future research direction.

# Declarations

## Ethical approval and consent to participate

Not applicable.

## Consent for publication

Not applicable.

## Data availability

The data and code can be obtained at: https://github.com/Mnb66/pLMFPPred.

## Competing interests

All authors claim no benefit conflicts or personal relationships that could have appeared to influence the work reported in this paper.

## Funding

This work was primarily supported by the National Natural Science Foundation of China (No.82270143) and Basic and Applied Basic Research Foundation of Guangdong Province, China (No.2022A1515220122).

## Authors' contributions

YZ conceptualized the study. YZ and JQH performed the analysis.WJY supervised the study. ZBM and XBH designed the model and conducted the experiments, ZBM and YLZ wrote this paper. HX and JXY provided suggestions and revised the manuscript. All authors read and approved the final manuscript.

## Acknowledgments

Not applicable.